\documentclass[twocolumn,showpacs,preprintnumbers,amsmath,amssymb]{revtex4}
\usepackage{dcolumn}
\usepackage{bm}
\usepackage[dvipdf]{graphicx}
\DeclareGraphicsExtensions{.jpg,.pdf,.mps,.png,.eps,.ps,.EPS}
			   
\begin{document}
\def\be{\begin{equation}}
\def\ee{\end{equation}}
\def\bc{\begin{center}}
\def\ec{\end{center}}
\def\bea{\begin{eqnarray}}
\def\eea{\end{eqnarray}}
\def\mx{{\mathbf x}}

\title{Detecting communities in large networks}
\author{A. Capocci$^1$, V.D.P. Servedio$^1$$^2$,
G. Caldarelli$^1$$^2$, F. Colaiori$^2$
}
\affiliation{$^1$Centro Studi e Ricerche e Museo della Fisica
  ``E. Fermi'', Compendio Viminale, Roma, Italy\\ 
$^2$INFM UdR Roma1-Dipartimento di Fisica Universit\`a ``La
  Sapienza'', P.le A. Moro 5, 00185 Roma, Italy} 
\begin{abstract}
We develop an algorithm to detect community structure in complex
networks. The algorithm is based on spectral methods and takes into
account weights and links orientations. Since the method detects
efficiently clustered nodes in large networks even when these are not
sharply partitioned, it turns to be specially suitable to the
analysis of social and information networks. 
We test the algorithm on a large-scale data-set from a psychological
experiment of word association. In this case, it proves to be
successful both in clustering words, and in uncovering mental
association patterns. 
\end{abstract}
\maketitle

Measurements and exact results concerning the clustering patterns of
networks mainly concern the occurrence of regular motifs
\cite{RMP,DoroRev,Eckmann02,Loops} and their correlations
\cite{Fitness03,Capocci03,Guido_cycles}. 
However, many social and information networks, such as the World Wide
Web, turn out to be approximately partitioned into communities of
irregular shape: for example, web pages focusing on similar topics are
strongly mutually connected and have a weaker linkage to the rest of the Web. 
The design of methods to partition a graph into several meaningful
highly inter-connected components have then become a compelling
application of graph theory to biological, social and information networks
\cite{Simonsen03,Kumar99,NewmanGirvan,NewmanRev03}.

Detecting the community structure in information networks allows one
to mine information in a more efficient way, narrowing the exploration
of a network as large as the World Wide Web (about $10^9$ nodes) to a
limited portion of it. 
When used in the analysis of large collaboration networks, such as
company or universities, communities reveal the informal organization
and the nature of information flows through the whole system
\cite{Huberman03,Guimera03}. 

There are several empirical methods to detect communities.  The
most successful algorithm, recently introduced by \cite{NewmanGirvan}
(NG--algorithm), is based on the edge betweenness, that measures the
fraction of all shortest paths passing on a given link, or,
alternatively, the probability that a random walk on the network runs
over that link. By removing links with high betweenness, one
progressively splits the whole network into disconnected components,
until the network is decomposed in communities consisting of one
single node. The outcome of the algorithm is represented by a
dendrogram, i.e. a tree--like diagram where each branching corresponds
to a splitting event. Though this method has been shown to be very
powerful in cases where some a priori knowledge of the a community
structure is given, it has two main disadvantages: first, that it
does not give an indication of the resolution of the clustering, and
thus it needs extra information as input (like the expected number of
clusters); second, that its outcome is independent on how sharp the
partitioning of the graph is. In the same spirit, \cite{Castellano03}
proposed an algorithm based on local analogues of the edge
betweenness. This has the advantage of being faster, but has the same
drawbacks on the NG--algorithm.

An alternative way to tackle the problem, which is the one we
pursue, is by spectral analysis. Previous approaches to graph
partitioning from spectral analysis have been mostly developed in the
computer science community to the purpose of finding the best
allocation of processes on processors in parallel computers, and are
based on iterative bisection. When applied to find communities
structures these methods have the disadvantage that repeated bisection
is not guaranteed to reach the best or most natural partition in
general cases. Moreover, they suffer from the same limitation of the
algorithm based on the edge betweenness, since they give no indication
of when the bisection should terminate, and thus need extra
information on the expected number of communities.

Our aim in this paper is to develop some spectral based algorithm 
able to reveal the structure of a complex network, which could be blurred
by the bias artificially over-imposed by the iterative bisection constraint.
Such a method should be able to conjugate the power of spectral analysis
to the caution needed to reveal an underlying structure when there is no 
clear cut partitioning, as is often the case in real networks. 


Spectral methods are based on the analysis of the adjacency matrix
 $A$ \cite{Hall70,Seary95,Kleinberg99}, whose element $a_{ij}$ is
 equal to $1$ if $i$ points to $j$ and $0$ otherwise. 
In particular, such methods analyze simple functions of $A$:
the Laplacian matrix $L=K-A$ and the Normal matrix $N = K^{-1}A$,
where $K$ is the diagonal matrix with elements $k_{ii}=\sum_{j=1}^S
a_{ij}$ and $S$ is the number of nodes in the network. 
In most approaches, referring to undirected networks, $A$ is assumed
 to be symmetric. 

The matrix $N$ has always the largest eigenvalue equal to one,
associated to a trivial constant eigenvector, due to row
normalization.  In a network with an apparent cluster structure, $N$
has also a certain number $m-1$ of eigenvalues close to one, where $m$
is the number of well defined communities, the remaining eigenvalues
lying a gap away from one. The eigenvectors associated to these
first $m-1$ nontrivial eigenvalues, also have a characteristic
structure: the components corresponding to nodes within the same
cluster have very similar values $x_i$, so that, as long as the
partition is sufficiently sharp, the profile of each
eigenvector, sorted by components, is step--like. The number of steps
in the profile corresponds again the number $m$ of communities.  A
similar information is encoded in the non-negative definite Laplacian
matrix, where the eigenvalues close to zero are associated to clusters.

The study of the eigenvectors profiles and the eigenvalues has
practical use only when a clear partition exists, which is rarely the
case. In most common occurrences, the number of nodes is too large and
the separation between the different communities is rather smooth.
Thus communities cannot be simply detected by looking at the first 
nontrivial eigenvector. 
We resolve this issue by combining information from the first few 
eigenvectors, and extracting the community structure from correlations 
between the same components in different eigenvectors. 

To describe the method in detail and understand why it works, it is
instructive to recast the eigenproblem into an optimization
problem. With the most general applications in mind, instead of the
adjacency matrix $A$, we focus on the weight matrix $W$, whose
elements $w_{ij}$ are assigned the intensity of the link $(i,j)$. We 
consider undirected graphs first, and then we pass to the most 
general directed case. 
Consider the following constrained optimization problem: 
Let $z(\mx)$ be defined as
\be
z(\mx) = \frac{1}{2}\sum_{i,j=1}^{S} (x_i - x_j)^2 w_{ij} \,,
\label{z}
\ee
where $x_i$ are values assigned to the nodes, with some constraint 
on the vector $\mx$, expressed by 
\be
\sum_{i,j=1}^{S} x_i x_j m_{ij} = 1 \,,
\label{constraint}
\ee
where $m_{ij}$ are elements of a given symmetric matrix $M$. 

The stationary points of $z$ over all $\mx$ subject to the constraint
(\ref{constraint}) are the solutions of 
\be (D-W) \mx = \mu M
\mx \,, 
\ee 
where $D$ is the diagonal matrix $d_{ij} =\delta_{ij}
\sum_{k=1}^S w_{ik}$, and $\mu$ is a Lagrange multiplier.

Different choices of the constraint $M$ leads to different eigenvalues
problems: for example choosing $M=D$ leads the eigenvalues
problem $D^{-1}W \mx = (1-2\mu) \mx$, while $M=1$ leads to
$(D-W)\mx=\mu \mx$. Thus $M=D$ and $M=1$, corresponds to the 
eigenproblems for the (generalized) Normal and Laplacian 
matrix respectively.

Thus, solving the eigenproblem is equivalent to minimizing the
function (\ref{z}) with the constraint (\ref{constraint}), were the
$x_i$'s are eigenvectors components. The absolute minimum corresponds
to the trivial eigenvector, which is constant. The other stationary points
correspond to eigenvectors where components associated to well
connected nodes assume similar values.

In order to compute cluster sizes and distribution, methods such as
bisection or edge-betweenness based ones are very poor in detect
the end of the recursive splitting. Our approach, instead, immediately
detects the number of clear clusters from the eigenvectors profile.

\begin{figure}
\includegraphics[width = 88 mm]{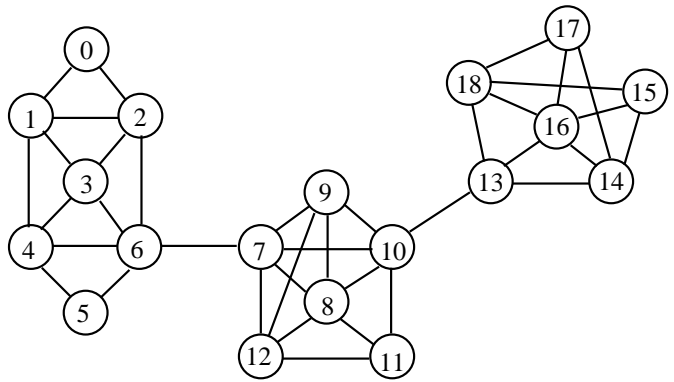} 
\caption{
Network employed as an example, with $S=19$ and random weights between $1$ and 
$10$ assigned to the links. Three clear clusters appear,
composed by nodes $0-6$, $7-12$ and $13-19$.
\label{toy}}
\end{figure}

\begin{figure}
\includegraphics[width = 88 mm, height = 60 mm]{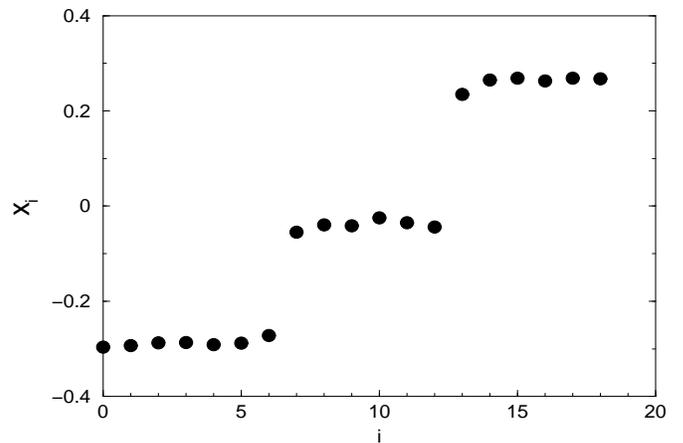} 
\caption{
Values of the 2nd eigenvector components for matrix $D^{-1} W$
relative to the graph depicted in figure \ref{toy}. 
\label{3clusters}}
\end{figure}

As an illustrative example, we show in Fig.\ref{3clusters} the
profile of the second eigenvectors of $D^{-1}W $ corresponding to the
simple graph shown in Fig.\ref{toy} with $S=19$ nodes, where random
weights between $1$ and $10$ were assigned to the links. The
components of the eigenvectors assume approximatively constant values
on nodes belonging to the same community. 
Thus, the number of communities emerges naturally and it is not
needed as input, .

However, as aforementioned, when dealing with large networks with no clear
partitioning, the precise value of the eigenvector components is
of little use. 
In such situations, the typical eigenvector profile is not step-like,
but resembles a continuous curve.   
Nevertheless, our method can still be applied, and efficiently
detects sets of well connected nodes.
In fact, components corresponding to nodes belonging to the same
communities are still strongly correlated taking, in each eigenvector, 
similar values among themselves.
Thus, a natural way to identify communities in an automatic manner, is
by measuring the correlation 
\begin{equation}
r_{ij} = \frac{ \langle x_i x_j \rangle -
\langle x_i \rangle \langle x_j \rangle }
{[(\langle x_i^2 \rangle - \langle x_i \rangle^2)
(\langle x_j^2 \rangle - \langle x_j \rangle^2)]^\frac{1}{2}}\,,
\end{equation}
where the average $\langle \cdot \rangle$ is over the first 
few nontrivial eigenvectors. 
The quantity $r_{ij}$ measures the community closeness between node
$i$ and $j$. Though the performance may be improved by averaging over 
more and more eigenvectors, with increased computational effort, 
we find that indeed a small number of eigenvectors suffices to identify 
the community to which nodes belong, even in large networks.

When dealing with a directed network, links do not correspond to any
equivalence relation. Rather, pointing to common neighbors is a
significant relation, as suggested in the sociologists' literature 
where this quantity measures the so-called {\em structural
  equivalence} of nodes \cite{Newman03}. Accordingly, in a directed
network, clusters should be composed by nodes pointing to a high
number of common neighbors, no matter their direct linkage. 
For directed networks, we thus modify our method in the streamline of
the HITS algorithm \cite{Kleinberg99}. The HITS algorithm was proposed 
on empirical bases to find the main communities in large oriented 
networks. It assumes that the largest components (in the absolute value) 
of eigenvectors of the matrices  $AA^T$ and $A^TA$ correspond to highly 
clustered nodes belonging to a single community. 
Such algorithm efficiently detects the main communities, even when 
these are not sharply defined. However, it becomes computationally 
heavy when one is interested in minor communities, which correspond 
to smaller eigenvalues. As explained in the undirected case, 
we tackle this issue by combining information from the first few 
eigenvectors, and extracting the community structure from correlations 
between the same components in different eigenvectors.

To detect the community structure in a directed network, we therefore
replace, in the previous analysis, the matrix $W$ with a matrix
$Y=WW^T$.  
This corresponds to replacing the directed network with an
undirected weighted network, where nodes pointing to common neighbors
are connected by a link, whose intensity is proportional to the total sum of
the weights of the links pointing from the two original nodes to the
common neighbors.
Then, one performs the analysis on the undirected network as described
previously. 
Thus, the function to minimize in this case is 

\begin{equation} 
y(\mx) = \sum_{ijl}^{1,S} (x_i - x_j)^2 w_{il} w_{jl} \,. 
\label{y}
\end{equation}
Defining  $Q$ as the diagonal matrix 
$q_{ij}=\delta_{ij}\sum_{lj=1}^{S} w_{il}w_{jl}$, 
the eigenvalue problem for the analogous of the generalized 
normal matrix, 
\be
Q^{-1} Y{\mx} = \lambda {\mx}
\ee
is equivalent to minimizing the function 
(\ref{y}) under the constraint  
$\sum_{ijl=1}^{S} x_i x_j q_{ij}=1$.

Tested on simple examples of directed networks, the algorithm
associated to the minimization of $y$, outperforms the one based 
on the minimization of $z$.

To test this spectral correlation-based community
detection method on a real complex network, we apply the 
algorithm to data from a psychological experiment reported in
reference \cite{Tenenbaum03}.
Volunteering participants to the research had to respond quickly by
freely associating a word (response) to another word given as input
(stimulus), extracted by a fixed subset.
Scientists conducting the research have recorded all the stimuli
and the associated responses, along with the
occurrence of each association. In the same spirit of past works \cite{Fontoura03}, 
we construct a network were words are nodes, and directed links are drawn 
from each stimulus to the corresponding responses, assuming that a 
link is oriented from the stimulus to the response. 
The resulting network includes $S=10616$ nodes, with an average
in-degree equal to about $7$. Taking into account the frequency of 
responses to a given stimulus, we construct the weighted adjacency 
matrix $W$. In this case, passing to the matrix $Y$ means 
that we expect stimuli giving rise to the same response to be
correlated.

The large-scale properties of semantic
\cite{Tenenbaum03,Sigman02,Dorogovtsev01} and syntactic
networks\cite{Ferrer03} corresponding to different languages have been
examined in past literature, mainly based on dictionaries and texts: a
strong similarity has emerged in such surveys, showing that
statistical features must refer to a common underlying structure
rather than to individual cultures. Interestingly, word graphs
studied so far are found to be complex networks, characterized by the
small world property and by power-law degree distribution
independently of the specific definition of the network \cite{Ferrer03b}.

The word association network is an ideal playground to test our
algorithm as, despite the large size of the networks, the
quality of clustering can be evaluated by a direct inspection 
to the yieldings. 
In large databases like this, were a partition in communities is not 
defined in a natural manner, there is no definite answer to what the 
best partition is. Rather, one is interested in finding groups 
of highly correlated nodes, or groups of nodes highly connected 
to a given one.  
Table \ref{communitiesAI} shows the most correlated words to three
test-words.
The correlation are computed by averaging over just 10 eigenvectors of
the matrix $Q^{-1} Y$: the results appear to be quite
satisfactory, already with this small number of eigenvectors. 

\begin{table} 
\begin{tabular}{|c|c|c|c|c|c|}
\hline
science & 1 & literature & 1 & piano & 1 \\ \hline 
scientific &  0.994 & dictionary &  0.994 & cello & 0.993 \\ \hline 
chemistry &  0.990 & editorial &  0.990 & fiddle & 0.992 \\ \hline 
physics &  0.988 & synopsis &  0.988 & viola & 0.990 \\ \hline 
concentrate &  0.973 & words &  0.987 & banjo & 0.988 \\ \hline 
thinking &  0.973 & grammar &  0.986 & saxophone & 0.985 \\ \hline 
test &  0.973 & adjective &  0.983 & director & 0.984 \\ \hline 
lab &  0.969 & chapter &  0.982 & violin & 0.983 \\ \hline 
brain &  0.965 & prose &  0.979 & clarinet & 0.983 \\ \hline 
equation & 0.963 & topic &  0.976 & oboe & 0.983 \\ \hline 
examine & 0.962 & English &  0.975 & theater & 0.982 \\ \hline 
\end{tabular}
\caption{The words most correlated to {\it science},
{\it literature} and {\it piano} in the eigenvectors of 
$Q^{-1} W W^T$. Values indicate the correlation.
\label{communitiesAI}
} 
\end{table}
Besides the performance in finding clusters of correlated words, 
our results are suggestive of the criteria according to which
the participants to the experiment have associated words.
As we observed, free associations are made by synonymy or antinomy,
syntactic role, and even by analogous sensory perception.

In conclusion, we have introduced a new method to detect communities
of highly connected nodes within a network. The method is based on
spectral analysis and takes into account the presence of weighted
links between nodes. Unlike previous spectral approaches, our method is not
based on iterative bisection. We have tested our algorithm on a real
network instance, built upon the records of a psychological
experiments. The algorithm proves to be successful in clustering nodes
(in this case, words) according to reasonable criteria, and provides
an automatic way to extract the most connected sets of nodes to a
given one in a set of over $10^4$. Given the broad range of
applicability, such method suggests a reliable way of clustering
large-scale networks occurring in different fields, including biology,
computer science and sociology. 

We enjoyed useful discussion and suggestions by Ramon Ferrer i Cancho and
Miguel-Angel Mu\~{n}oz.

We acknowledge partial support from the FET Open Project IST-2001-33555 COSIN.


\end{document}